\documentclass[english,twocolumn,showpacs,aps,superscriptaddress,amsmath,amssymb,10pt,showkeys,prb]{revtex4-1}

\usepackage[utf8]{inputenc}
\usepackage{amsfonts}
\usepackage{graphicx}
\usepackage{subfigure}
\usepackage{appendix}
\usepackage{dcolumn}% Align table columns on decimal point
\usepackage{bm} % bold math
\usepackage[normalem]{ulem}
\usepackage[pdftex,usenames,dvipsnames]{color}
\usepackage{hyperref}

\usepackage{amsmath}

\allowdisplaybreaks
\makeatletter
\newcommand*{\supplementary}{%
  \close@column@grid
  \clearpage
  \twocolumngrid
}
\makeatother
\newcommand{\norm}[1]{\left\lVert#1\right\rVert}
\def\eqref#1{(\ref{#1})}
\def\angstrom{{\mbox{\AA}}}
\newcommand{\un}[1]{\,\mathrm{#1}}
\newcommand\smallO{
  \mathchoice
    {{\scriptstyle\mathcal{O}}}% \displaystyle
    {{\scriptstyle\mathcal{O}}}% \textstyle
    {{\scriptscriptstyle\mathcal{O}}}% \scriptstyle
    {\scalebox{.7}{$\scriptscriptstyle\mathcal{O}$}}%\scriptscriptstyle
  }

\definecolor{tangerine}{rgb}{0.944,0.522,0}
\definecolor{verde}{rgb}{0.,0.6,0}
\definecolor{rosso}{rgb}{0.9,0.0,0.2}
\definecolor{blue}{rgb}{0.0,0.0,0.8}

\newcommand{\editor}[2]{%
  \expandafter\newcommand\csname #1note\endcsname[1]{%
    \textcolor{#2}{(\textbf{#1:} \textit{\small ##1})}}%
  \expandafter\newcommand\csname #1\endcsname[1]{%
    \textcolor{#2}{##1}}%
  \expandafter\newcommand\csname #1cancel\endcsname[1]{%
    \textcolor{#2}{\sout{##1}}}%
  \expandafter\newcommand\csname #1change\endcsname[2]{%
    \textcolor{#2}{\sout{##1} ##2}}%
  \newenvironment{#1text}{\color{#2}}{\color{black}}
}

\editor{SB}{tangerine}
\editor{LE}{blue}
\editor{AM}{verde}
\editor{CORR}{black}
\begin{document}

\author{Aris Marcolongo}
\email{ISR@zurich.ibm.com}
\affiliation{Cognitive Computing and Computational Sciences Department, IBM Research – Z\"{u}rich, S\"{a}umerstrasse 4, CH-8803 R\"{u}schlikon, Switzerland}
\author{Loris Ercole}
\affiliation{SISSA -- Scuola Internazionale Superiore di Studi Avanzati, Via Bonomea 265, 34136 Trieste, Italy}
\altaffiliation{(Present address:) Theory and Simulation of Materials (THEOS), and National Centre for Computational Design and Discovery of Novel Materials (MARVEL), \'Ecole Polytechnique F\'ed\'erale de Lausanne, CH-1015 Lausanne, Switzerland.}
\affiliation{SISSA -- Scuola Internazionale Superiore di Studi Avanzati, Via Bonomea 265, 34136 Trieste, Italy}
\author{Stefano Baroni} 
\affiliation{SISSA -- Scuola Internazionale Superiore di Studi Avanzati, Via Bonomea 265, 34136 Trieste, Italy}
\altaffiliation{CNR -- Istituto Officina dei Materiali, SISSA, 34136 Trieste}

\keywords{ %Use showkeys class option if keyword display desired
  Transport properties,
  Molecular dynamics,
  Thermal conductivity,
  Statistical analysis of time series,
  Gauge invariance of transport coefficients \\
  PACS NUMBERS: 05.60.-k % Transport processes
  05.45.Tp % Time series analysis
  66.10.cd
}

%\SectionNumbersOn

%ACTIVATE FOR ACHEMSO
%\begin{document}

%REMOVE FOR ACHEMSO
\title{Gauge fixing for heat-transport simulations}
%\maketitle

\begin{abstract}
Thermal and other transport coefficients were recently shown to be largely independent of the microscopic representation of the energy (current) densities or, more generally, of the relevant conserved densities/currents. In this paper we show how this \emph{gauge invariance}, which is intimately related to the intrinsic indeterminacy of the energy of individual atoms in interacting systems, can be exploited to optimize the statistical properties of the current time series from which the transport coefficients are evaluated. To this end, we introduce and exploit a variational principle that relies on the metric properties of the conserved currents, treated as elements of an abstract linear space. Different metrics would result in different variational principles. In particular, we show that a recently proposed data-analysis technique based on the theory of transport in multi-component systems can be recovered by a suitable choice of this metric. 
\end{abstract}

\maketitle

\section{Introduction}

The concept of \emph{gauge invariance} of transport coefficients was introduced and explored in recent works on adiabatic heat \cite{Marcolongo2016,Ercole2016} and charge transport \cite{Grasselli2019} in electronic insulators. In its generality, this principle asserts that transport coefficients, such as thermal and electrical conductivities, are largely independent of the detailed form of the local representation of the conserved quantity (energy, charge, mass) being transported: any two such representations resulting in the same integrated value of the conserved quantity, and whose space correlations are short-range, are bound to yield the same transport coefficient. This finding was instrumental in establishing a rigorous and practicable density-functional theory (DFT) of adiabatic heat transport,\cite{Marcolongo2016,Baroni2018} based on the Green-Kubo (GK) linear-response approach,\cite{Green1952,*Green1954,Kubo1957a,*Kubo1957b} and more recently spurred new applications based on different definitions of the heat current.\cite{Kang2017,English2017,*Tse2018}
On the other hand, the question naturally arises as to how to exploit this \emph{gauge freedom} in order to optimize the statistical properties of the current time series from which transport coefficients are evaluated, and thus minimize the statistical errors affecting the latter.

\color{black}
 This work is devoted to the analysis of possible gauges for thermal transport simulations, i.e. different equivalent microscopic definitions of the heat flux, within the GK framework. The GK approach exploits the fluctuation-dissipation principle, which permits to evaluate non-equilibrium transport coefficients by analyzing the fluctuations of the heat flux during equilibrium molecular dynamics (MD) simulations. 
 We note that other frameworks have been successfully applied to evaluate the thermal conductivity based on MD simulations as well, but requiring the direct simulation of non-equilibrium transport processes. 
 We recall here for example the \emph{approach to equilibrium} technique,\cite{LAMPIN-2012} based on a transient, non stationary, process, and the M\"{u}ller-Plathe method,\cite{MP-1997} imposing an external (kinetic) energy flux by swapping particle velocities. These approaches circumvent the problem of definition of an energy density but require simulation sizes large enough to define a local temperature, and may be affected by non-linear effects. Nevertheless, they have been successfully applied in first-principles and classical frameworks.\cite{Puligheddu2017,Stackhouse2010thermal,Melis-AEMD-2014,Zhang-MP-2005} We cite for completeness an other class of techniques to evaluate thermal conductivity which avoids performing any MD simulation. These methods start with a lattice picture and introduce anharmonic effects via the mesoscopic Boltzmann transport equation. We refer the interested reader to a recent review \cite{Lind2019} and references therein.

The GK approach to thermal conductivity, developed much before non-equilibrium approaches, is still an area under active research. Efforts to improve the GK framework are dedicated 
to estimate size and simulation times required for the evaluation of the transport coefficient\cite{Ercole2017,Oliveira2017}. Other lines of research aim at developing novel definitions of the energy flux. 
For example, Ref.~\citenum{Kang2017} introduces a mask function to deal with periodic boundary conditions and an \emph{ad hoc} definition of atomic energies from first principles; Ref.~\citenum{CARBOGNO-2017} develops a first-principles heat flux where convective components are neglected; Ref.~\citenum{English2017,*Tse2018} uses an energy-moment perspective based on the Einstein formulation to estimate the thermal conductivity of solids.

The goal of this work is twofold. First, we want to highlight how the presence of spurious signals can hinder the convergence of GK approaches, especially in first-principles calculations of thermal conductivity of molecular systems.
Second, we introduce a general variational approach to fix the transport gauge so as to optimize the statistical properties of the estimator of the conductivity, thus reducing simulation times as much as possible. 
\color{black}
Such an operation reveals to be beneficial in keeping the statistical noise of the conductivities estimated from first-principles simulations at an acceptable level.
\color{black}
We stress that all the methodologies proposed in this work are based on the GK formalism and aim at modifying the heat fluxes used in the GK formulas in a computationally easy but effective way.
From all other aspects, the GK formalism remains unchanged.
\color{black}

\CORR{For most of the work, we} focus on heat transport in solids and one-component, possibly molecular, liquids, for which energy is the only relevant conserved quantity. We thus neglect multi-component liquid systems, where the total momenta of individual species should be considered to correctly define the heat current \cite{Baroni2018,Bertossa2019,Salanne_2011}. \CORR{Nevertheless, we will also discuss a theoretical relation between our formalism and the GK formulas derived in a multi-component setting.}
When dealing with quantum simulations,
we further assume the materials to be electronic insulators, thus excluding electrons as heat carriers.
Our discussion will be otherwise as general as possible and will include examples from classical as well as \emph{ab initio} molecular dynamics (MD). For the latter, we will follow the GK formulation developed in Ref.~\citenum{Marcolongo2016}.

The rest of the work is organized as follows. 
In Sec.~\ref{sec:theory} we introduce our general variational approach and discuss how it can be used to eliminate the ineffective but slowly-decaying signals that may appear in the classical and first-principles simulation of transport coefficients.
In Sec.~\ref{sec:techniques} we present two different implementations of the general variational principle, and
Sec.~\ref{sec:comparison} is devoted to their comparison.
In Sec.~\ref{sec:multi-component} we show how a recently introduced data-analysis technique, based on the theory of multi-component systems,\cite{Bertossa2019} can be derived from the variational principle outlined in the present paper, thus
providing novel insight into the known formulas. 
Finally, Sec.~\ref{sec:conclusions} contains our conclusions.

\section{Theoretical Background}  \label{sec:theory}

\subsection{Variational formulation} \label{sec:theory-variational}

In one-component isotropic systems the GK equation expresses the thermal conductivity in terms of the time autocorrelation function of the energy flux as:
\begin{equation}
    \kappa = \frac{1}{3Vk_B T^2} \int_0^\infty \! \left\langle \mathbf{J}(t) \cdot \mathbf{J}(0) \right\rangle dt , \label{eq:GK}
\end{equation}

where $V$ is the system's volume, $k_B$ the Boltzmann constant, $T$ the temperature, $\mathbf{J}(t)=\int\mathbf{j}(\mathbf{r},t) d\mathbf{r}$ is the macroscopic energy current, usually referred to as the \emph{heat} (or energy) \emph{flux}, $\mathbf{j}(\mathbf{r},t)$ is the energy current density and $\langle\cdot\rangle$ indicates an equilibrium average over molecular trajectories in the microcanonical (NVE) ensemble. \cite{Green1952,*Green1954,Kubo1957a,*Kubo1957b,Esfarjani2011,stackhouse2010,Baroni2018} The integral of the autocorrelation function of the energy flux in Eq.~\eqref{eq:GK} can also be interpreted as the zero-frequency component of its power spectrum, and is often called the GK integral. The energy flux $\mathbf{J}(t)$ is a classical observable, \emph{i.e.} a function defined over the phase space of the system: $\mathbf J(t) = \mathbf J(\Gamma_t)$, where $\Gamma_t\equiv \{\mathbf r _i(t), \mathbf v _i (t) \}$ indicates the system's phase-space position at time $t$ and $\mathbf r_i$ and $\mathbf v_i$ the position and velocity of the $i$-th atom. The set $\mathcal O$ of all such observables is naturally endowed with the structure of a real vector space, via pointwise addition and multiplication. 

In order to proceed further, we first define an \emph{inert} or \emph{non-diffusive} flux as one such that its GK integral, Eq.~\eqref{eq:GK}, vanishes. Interestingly, adding a non-diffusive energy flux to a diffusive one does not change the value of the conductivity calculated from the latter, an intuitive result that was rigorously proven in Ref.~\citenum{Marcolongo2016} and that is at the basis of the whole concept of gauge invariance of transport coefficients. Nevertheless, non-diffusive terms may increase the power of the noise of the time series to such a level as to compromise its analysis and thus making a numerical estimate of the conductivity too expensive.
This suggests one to devise new ``optimized'' definitions of the energy flux, whereby signal components known to be non-diffusive are conveniently eliminated, thus reducing the power of the noise without affecting the value of the signal. 

Let us suppose for the moment that a set of linearly independent such non-diffusive signals $\lbrace\mathbf{Y}_n\rbrace$, $n=1,\dots N$ can be identified. Then, the new time series
 \begin{equation}
    \mathbf{J}' \equiv \mathbf{J} - \sum_n \lambda_n \mathbf{Y}_n  \label{eq:new-current},
\end{equation}
will provide, \CORR{by replacing $\mathbf{J}$ with $\mathbf{J}'$ into Eq.~\eqref{eq:GK} and} with infinite sampling, the same thermal conductivity as the one computed using the original heat flux $\mathbf{J}$ for any choice of the $\lbrace\lambda_n\rbrace$ coefficients. In order to minimize the statistical noise, we should minimize the magnitude of inert signals. To achieve this goal, let us introduce a generic scalar product $(\cdot,\cdot)$ on the vector space $\mathcal O$ of observables, inducing a norm $\norm{A} \equiv \sqrt{(A,A)}$. For the moment, we leave the scalar product unspecified. We now propose to choose as optimal coefficients those that minimize the function:
\begin{equation}
   f(\lbrace\lambda_n\rbrace) \equiv \norm{ \mathbf{J} - \sum_{n} \lambda_n \mathbf{Y}_n }^2    \label{eq:function},
\end{equation}
which is the standard loss function used in a multiple-linear regression framework. From a geometrical point of view, the chosen $\mathbf{J}'$ is the component of $\mathbf{J}$ orthogonal to the subspace generated by the non-diffusive signals. We note that, as long as the scalar product is positive definite (\emph{i.e.} $\left( A, A \right) > 0$, if $A \ne 0$) and the non-diffusive signals are linearly independent the quadratic form in Eq.~\eqref{eq:function} is positive definite and has a unique minimum.

\CORR{Finally, we note also that the variational framework applies directly to the case of anisotropic systems as well. Thermal conductivity becomes then a tensor $\kappa_{\alpha,\beta}$, a function of the cross correlations $\left\langle J^{\alpha}(t) J^{\beta}(0) \right\rangle$. In this case it is possible 
to optimize $\mathbf{J}'$ component by component: $J'_{\alpha}=J_{\alpha}-\sum_n \lambda_{n,\alpha} Y_{n,\alpha}$, where the $\lambda_{n,\alpha}$ are chosen to minimize $\norm{J_{\alpha}-\sum_n \lambda_{n,\alpha} Y_{n,\alpha}}^2$, for each $\alpha$ under the chosen norm. Evaluating the thermal conductivity tensor using $\mathbf{J}'$ instead of $\mathbf{J}$ in the tensorial version of Eq.~\eqref{eq:GK} would still lead to the same thermal conductivity thanks to the non-diffusivity of the signals $Y_{n,\alpha}$, but this replacement is expected to reduce the computational cost.
For simplicity, in this work we focus on isotropic systems. 
}

\subsection{Inert signals in \emph{ab initio} simulations} \label{sec:slow-decay}
The estimate of thermal conductivity from the GK equation, \eqref{eq:GK}, poses serious numerical challenges, especially in \emph{ab initio} MD simulations, where the size and time scales that can be afforded are limited. Even though efficient and automated estimators of $\kappa$ have been recently devised,\cite{Ercole2017} non-diffusive components may increase the variance of the energy-flux time series considerably, thus slowing down the convergence of the estimator, as for example it was observed in the case of liquid water \cite{Marcolongo2016} or silica glass.\cite{Ercole2018}

One of the most important sources of inert signals that can be identified in \emph{ab initio} simulations is related to the large atomic binding energies affecting the Born-Oppenheimer energies of extended systems. In order to understand what is at stake here, let us consider the case where an all-electron picture of the electronic structure is adopted. Evidently, the nuclei undergoing thermal diffusion drag core electrons around and, with them, the large binding energies rigidly attached to them: these energies are not available to any physical processes accessible at thermal energies and do not contribute therefore to heat transport, while they do increase the magnitude of the energy-flux fluctuations, thus enormously enhancing the noise affecting the estimator of the conductivity. Even in a pseudo-potential picture, cohesion energies are a small fraction of the Born-Oppenheimer total energies, which are therefore dominated by the atomic \emph{binding energies}, $\{\epsilon^S\}$, defined as the energies necessary to remove all the valence electrons from an isolated atom of the $S$-th species. The energy density associated to such binding energies has the form: $\sum_i \epsilon^{S(i)} \delta (\mathbf{r}-\mathbf{r}_i)$, $S(i)$ being the species of the $i$-th atom, and $\mathbf{r}_i$ its position.

To fix the notation, in this work capital Latin letters refer to atomic species, and Latin lowercase letters refer to atomic indexes; Greek letters are reserved for Cartesian coordinates, and a summation over repeated Cartesian indexes is implied. We denote with $\mathbf V^{S} =\sum_{\{i\in S\}} \mathbf{v}_i$ the sum of the velocities of all the atoms of species $S$, which is sometimes called the macroscopic \emph{particle current}. We call $\mathbf{W}^S$ the velocity of the center of mass of the corresponding atomic species: $\mathbf{W}^S = \mathbf{V}^S / N^S$, where $N^S$ is the number of atoms of species $S$. According to these definitions, the energy current associated to the binding energy reads:
\begin{equation}
    \mathbf{J}^\mathrm{bind}=\sum_S N^S \epsilon^S \mathbf W^S.  \label{eq:bind-flux}
\end{equation}
In a mono-atomic fluid, $\mathbf{J}^\mathrm{bind}$ is constant because of momentum conservation and it vanishes in the center-of-mass reference frame; in solids, it is clearly non-zero and non-diffusive, since atoms individually do not diffuse; and in molecular fluids it can be proven that the particle currents are non-diffusive as well.\cite{Marcolongo2014,Marcolongo2016,Baroni2018}

In general, the energy of isolated atoms is not considered in classical, semi-empirical potentials. Nevertheless, the time-averaged atomic energies formally behave as atomic binding energies and they can be safely subtracted from the definition of the atomic energies entering the expression of the energy flux, without altering the value of the resulting heat conductivity; \emph{i.e.} the heat conductivity is invariant with respect to the transformation $\epsilon_i\rightarrow \epsilon_i - \bar \epsilon^{S(i)}$, where $\bar \epsilon^{S}$ is the average energy of atoms of species $S$. In general, classical mean atomic energies are much smaller than the typical atomic binding energies in first-principles simulations, thus leading to smaller inert fluxes.

\section{Techniques}  \label{sec:techniques}

The most common choice of scalar product between two real observables $A, B \in \mathcal O$ is given by the static cross-correlation, which we refer to as the \emph{microcanonical (MC) scalar product}:
\begin{align}
    \left( A, B \right)_\mathrm{MC} &\equiv \langle A B \rangle = \int \! P_\mathrm{NVE}(\Gamma) A(\Gamma) B(\Gamma) \, d\Gamma ,  \label{eq:boltzmann-scalar-product}
\end{align}
where $P_\mathrm{NVE}$ is the constant energy distribution of the microcanonical ensample, explored ergodically via Hamiltonian dynamics. \color{black} For vector observables, the product $AB$ implies a scalar product as well \color{black}. The last expression makes it clear that the MC scalar product is positive definite. In this section we explore different optimization techniques based on this choice of the metric in Eq.~\eqref{eq:new-current} and \eqref{eq:function}.

The minimimization of Eq.~\eqref{eq:function}, \emph{i.e.} the projection onto the subspace orthogonal to the inert fluxes, can be directly performed by differentiating Eq.~\eqref{eq:function} with respect to $\lambda_{m}$ and obtaining the linear system:
\begin{equation}
    \langle \mathbf{J}\cdot \mathbf{Y}_n \rangle - \sum_{m} \lambda_{m} \langle \mathbf{Y}_{m} \cdot \mathbf{Y}_n \rangle = 0, \quad  n=1,\dots N.  \label{eq:decorr-LS}
\end{equation}
The thermal conductivity can be then computed via the GK equation \eqref{eq:GK}, by replacing $\mathbf{J}$ with $\mathbf{J}'$, that is defined in Eq.~\eqref{eq:new-current} using the $\lbrace\lambda_n\rbrace$ coefficients that solve Eq.~\eqref{eq:decorr-LS}. This technique has been instrumental in computing the heat conductivity of liquid water from an energy-flux time series that would have been otherwise affected by an intractable numerical noise.\cite{Marcolongo2016}
We note that the new energy current $\mathbf{J}'$ is computed from the difference of the original current and the weighted inert fluxes, \emph{i.e.} a difference of signals of large amplitude.
We refer to this procedure as the \emph{decorrelation technique}, since the solution of Eq.~\eqref{eq:decorr-LS} is equivalent to imposing that the current $\mathbf{J'}$ is decorrelated with respect to each inert signal $\lbrace\mathbf{Y}_n\rbrace$.

We now discuss how, when the $\mathbf Y$ currents are assumed to coincide with the average velocities of the various atomic species,  $\{\mathbf Y_n \}  = \{\mathbf{V}^S\}$, decorrelation is equivalent to a different \emph{renormalization} procedure of the individual atomic velocities. 
Velocity renormalization does not require the solution of any linear system and involves only operations with small-amplitude signals. Let $\mathbf{J}_{\mathrm{bare}} (\{\mathbf{r}_i, \mathbf{v}_i\})$ be the heat current, a function of the atomic coordinates $\{\mathbf{r}_i,\mathbf{v}_i\}$. The renormalized heat current is defined as:
\begin{equation}
    \mathbf{J}_{\mathrm{ren}}(\{\mathbf{r}_i,\mathbf{v}_i\}) \equiv \mathbf{J}_{\mathrm{bare}}(\{\mathbf{r}_i,\mathbf{v}_i-\mathbf{W}^{S(i)}\}) , \label{eq:RC-definition}
\end{equation}
where in the renormalized velocities $\mathbf{v}_i' \equiv \mathbf{v}_i - \mathbf{W}^{S(i)}$ the velocity of the centre of mass of the respective species has been subtracted. The renormalized current can be computed without any coding effort by any program providing an energy current by replacing input velocities with their renormalized values and leaving the positions unchanged. The non diffusive particle current $\mathbf{V}^S$ has been effectively set to zero in the definition of $\mathbf{J}_\mathrm{ren}$. 

We demonstrate the effectiveness of the renormalization procedure by performing several classical MD simulations of liquid water at ambient conditions, adding a fictitious formation energy term to each species and studying the effects on the resulting energy currents. Simulations were carried out using the \textsc{LAMMPS} molecular dynamics code,\cite{LAMMPS} at ambient conditions, \CORR{in a cubic box containing $216$ water molecules,} considering a flexible model of water \cite{water-FF} and an integration time step of $0.5\un{fs}$ \CORR{(further simulation details are reported in Appendix~\ref{app:details})}. 
A species-dependent shift was added to the instantaneous atomic energies: $\epsilon_i \rightarrow \epsilon_i + \overline{\epsilon}^{S(i)}$. We chose small values for the formation energies: $\overline{\epsilon}^{H}\approx -0.2, -0.4 \un{eV}$, and $\overline{\epsilon}^{O}\approx -0.4, -0.9 \un{eV}$, \emph{i.e.} values that are higher than the typical interaction energies but much lower than the formation energies considered in DFT. Even such a small perturbation can have an impact on the convergence properties of the GK current.
\begin{figure}[!tb]
    \centering
    \subfigure[\label{fig:RC-invariance-gk}]{\includegraphics[scale=1.0]{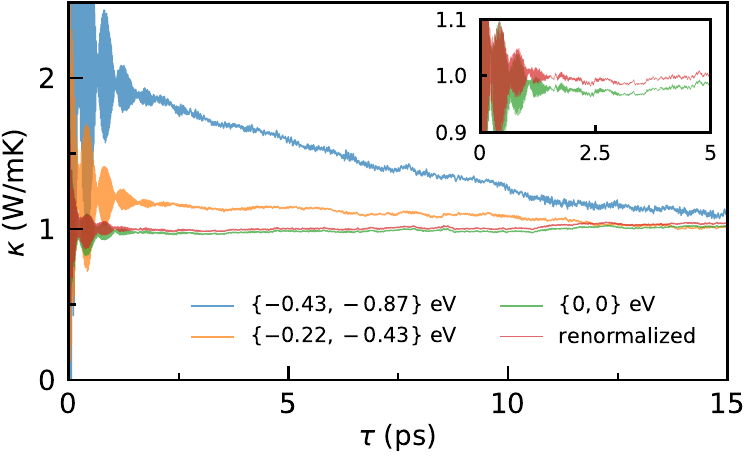}} \hfill
    \subfigure[\label{fig:RC-invariance-spectra}]{\includegraphics[scale=1.0]{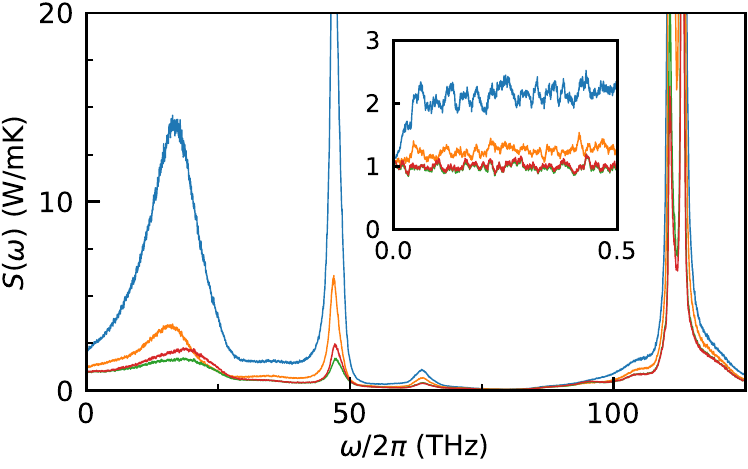}}
    \caption{
    Convergence behavior of the thermal conductivity obtained from different definitions of the energy current of classical liquid water. Species-dependent formation energies $\{\bar{\epsilon}^H,\bar{\epsilon}^O\}$ were added (see legend and text). In particular, the red line represents the renormalized current.
    (a) Thermal conductivity computed from the GK equation, Eq.~\eqref{eq:GK}, as a function of the upper integration limit 
    \CORR{(see Appendix~\ref{app:details} for a numerical definition). Inset: zoom of the low-$\tau$ region.}
    (b) Power spectrum of the energy current, whose zero-frequency value is the thermal conductivity. Inset: zoom of the low-frequency region.
    }
\end{figure}
In Fig.~\ref{fig:RC-invariance-gk} we plot the thermal conductivity estimated by a direct integration of the energy-flux autocorrelation function, Eq.~\eqref{eq:GK}, as a function of the upper integration limit \CORR{(see Appendix~\ref{app:details} for a numerical expression)}, using the original definition of the energy current ($\mathbf{J}_{\mathrm{bare}}^{\lbrace 0,\,0\rbrace}$), several definitions with additional formation energies ($\mathbf{J}_{\mathrm{bare}}^{\lbrace\bar{\epsilon}_\mathrm{H},\,\bar{\epsilon}_\mathrm{O}\rbrace}$), and the latter after renormalization was applied ($\mathbf{J}_\mathrm{ren}$). The addition of formation energies increases the variance of the time series and slows down the convergence of the GK integral, thus requiring one to run longer simulations in order to converge the integral with similar accuracy. Nevertheless, as expected, all the integrals converge to the same value after a sufficient integration time. Once the renormalization procedure is performed, the convergence becomes much faster and the integral resembles to the one computed from $\mathbf{J}_{\mathrm{bare}}^{\lbrace 0,\,0\rbrace}$. Notice that renormalizing any of the time series returns exactly the same $\mathbf{J}_\mathrm{ren}$. Additional information can be inferred from the power spectrum of the energy current, defined as $S(\omega) = \frac{1}{\mathcal{T}}\left\langle\left|\int_0^\mathcal{T} \mathbf{J}(t) \mathrm{e}^{i\omega t} dt \right|^2\right\rangle$ and plotted in Fig.~\ref{fig:RC-invariance-spectra}, which clearly shows that formation energies lead to an increase in the power of the signal (the integral of the spectrum). The zero-frequency value of the spectrum, which is proportional to the thermal conductivity,\cite{Ercole2017,Baroni2018} is not affected by the definition used, but becomes more and more difficult to estimate when larger formation energies are considered, due to the fast increase of $S(\omega)$ at $\omega\sim 0$.

\begin{figure}[!tb]
    \centering
    \subfigure[\label{fig:silica-gk}]{\includegraphics[scale=1.0]{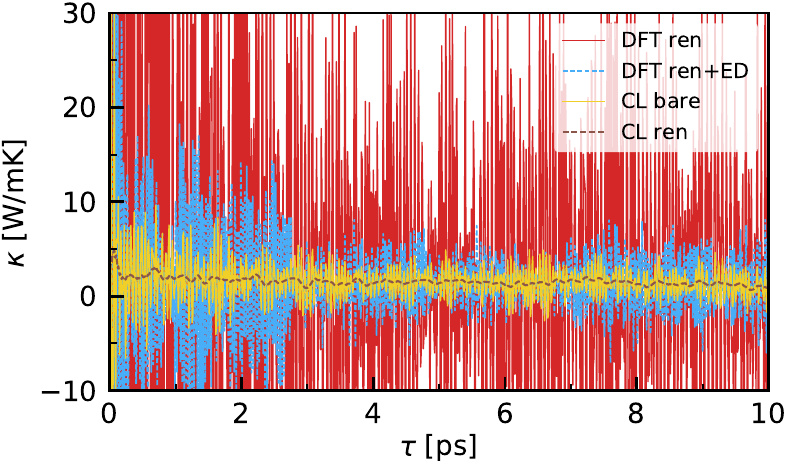}} \hfill
    \subfigure[\label{fig:silica-spectra}]{\includegraphics[scale=1.0]{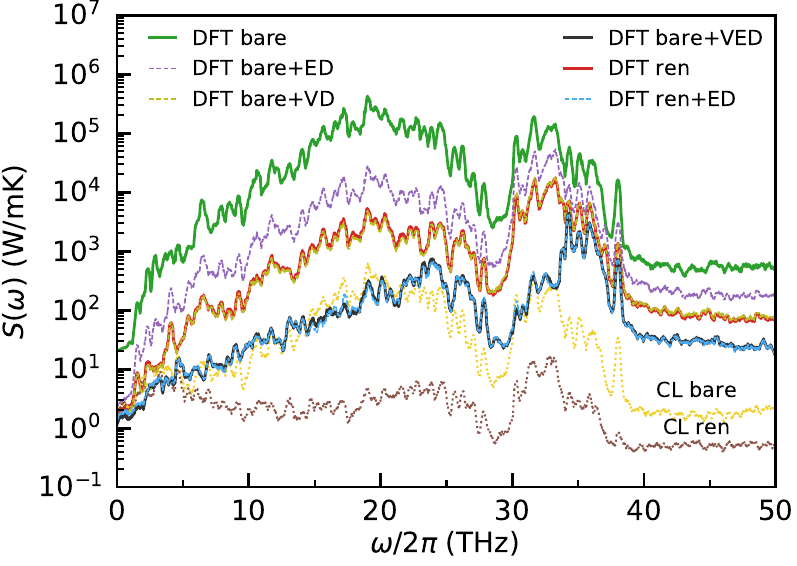}}
    \caption{
    Analysis of different definitions of the energy current of a 72-atom sample of a-SiO$_2$. The original $\mathbf{J}_\mathrm{bare}$ classical (CL) and DFT definitions have been renomalized, obtaining $\mathbf{J}_\mathrm{ren}$. Also reported are the same currents decorrelated with respect to the particle current of one species (VD), the electronic current (ED), or both (VED).
    (a) Thermal conductivity computed from the GK equation, Eq.~\eqref{eq:GK}, as a function of the upper integration limit. The result of the bare DFT current is out of scale and has been omitted. 
    (b) Power spectrum of the energy currents. The zero frequency corresponds to $\kappa$.
    }
\end{figure}

We now repeat the same kind of analysis in an \emph{ab initio} framework considering amorphous silica, a-SiO$_2$, a multi-component solid, at $\sim 350\un{K}$. 
\CORR{For reference, in Appendix~\ref{app:details} we report the formulas used to evaluate the bare first-principles energy current.}
For the purpose of this calculation and to save computational time we considered a small cell of $72$ atoms, we generated the trajectory via the classical BKS interatomic potential\cite{Silica-BKS-1990}, and for $10000$ snapshots (corresponding to $100\un{ps}$ of trajectory) we computed the energy current using the classical and \emph{ab initio} definitions \CORR{(further simulation details in Appendix~\ref{app:details})}. Finally, we applied the renormalization procedure to the latter. Let us remark that in order to compute $\kappa$ from first principles one should generate the trajectory by \emph{ab initio} MD, nevertheless we use this expedient to observe the effects of the different energy-current definitions, without including the effects of the different dynamics. For the DFT calculations we used the \textsc{Quantum ESPRESSO} package\cite{Giannozzi2017} with a PBE functional\cite{Perdew1996}, a plane-wave cutoff of $80\un{Ry}$, and optimized norm-conserving Vanderbilt pseudopotentials (ONCVP)\cite{Hamann2013,*Schlipf2015}. In Fig.~\ref{fig:silica-gk} we plot the GK integrals, which feature very large oscillations, making a direct estimate of $\kappa$ extremely difficult. In particular, the integral of the DFT bare energy current $\mathbf{J}^\mathrm{DFT}_\mathrm{bare}$ is out of scale and is not shown.  Much more information can be gained from the power spectra of the currents, reported in Fig.~\ref{fig:silica-spectra} on a logarithmic scale. We immediately notice the much larger power of the $\mathbf{J}^\mathrm{DFT}_\mathrm{bare}$ time series (labeled ``DFT bare'', top left) with respect to the classical one (labeled ``CL bare'', bottom right). Indeed, the standard deviation of $\mathbf{J}^\mathrm{DFT}_\mathrm{bare}$ (proportional to the integral of the spectrum) is $25$ times larger than the classical one $\mathbf{J}^\mathrm{cl}_\mathrm{bare}$, and in both cases the renormalization reduces the standard deviations of the currents by a factor of $\approx 5$. The huge power of the bare DFT current makes an estimate of $\kappa$ impossible to converge to physical values with the trajectory lengths attainable with \emph{ab initio} MD. Renormalization allows one to solve this problem and to obtain meaningful results, which are also compatible with the classical values, as reported in Table~\ref{tab:kappa-results} in the fields ``DFT ren'', ``Classic bare'', and ``Classic ren''. 
\CORR{Although these simulations were not meant to obtain quantitatively accurate values, the thermal conductivities are in fairly good agreement with previous experiments and computational studies ($\kappa_\mathrm{exp} \approx 1.4\un{W/mK}$)\cite{Yamane2002,*Larkin2014,*Touloukian1970,*Cahill1990}.
More reliable and accurate results would require a more careful choice of the force field and a study of the system-size dependence of $\kappa$, which is especially critical in the case of amorphous systems.\cite{Ercole2018}}

\begin{table}[!tb]
    \centering
    \begin{tabular}{l|cc}
        $\mathbf{J}$ definition &   a-SiO$_2$        &   H$_2$O \\
        \hline
        DFT bare       &   $25.5 \pm 5.9$   &   $740 \pm 140$ \\
        DFT ren        &   $1.31 \pm 0.29$  &   $1.18 \pm 0.17$ \\
        Classic bare   &   $1.32 \pm 0.29$  &   -- \\
        Classic ren    &   $1.28 \pm 0.26$  &   -- \\
        \hline
        DFT bare, VD   &   $1.30 \pm 0.28$   &   $0.99 \pm 0.15$ \\
        DFT bare, ED   &   $2.79 \pm 0.61$   &   $341 \pm 46$ \\
        DFT bare, VED  &   $0.95 \pm 0.21$   &   $0.82 \pm 0.12$ \\
        DFT ren, ED    &   $1.11 \pm 0.24$  &   $1.03 \pm 0.15$ \\
    \end{tabular}
    \caption{
    Thermal conductivities ($\mathrm{W/mK}$) estimated via cepstral analysis from different definitions of the energy current (parameters used in the analysis are reported in Appendix~\ref{app:details}). 
    VD, ED, and VED indicate a current that has been decorrelated with respect to the particle current of a species $\mathbf{V}^S$, the electronic current $\mathbf{J}_\mathrm{el}$, or both, respectively. Errors are one standard deviation. }
    \label{tab:kappa-results}
\end{table}

It is possible to prove formally the equivalence of the transport coefficients computed from the bare and renormalized currents, $\kappa_{\mathrm{bare}}=\kappa_{\mathrm{ren}}$, starting with the following generic expression of a component of a GK energy flux:
\begin{equation}
    J_{\mathrm{bare},\alpha} = \sum_{i}\left(\frac{1}{2}m_{i}v_i^2\right)v_{i,\alpha} + \sum_i \varepsilon^{i}_{\alpha\beta}(\{\mathbf{r}\})\, v_{i,\beta}. \label{eq:generic}
\end{equation} 
Any atomic energy current, defined as the time derivative of the first moment of an energy density, can be reduced to such a form, the matrices $\varepsilon^{i}_{\alpha\beta}$ being functions of coordinates only (dependent on the local environment) and not of particle velocities. 
From Eq.~\eqref{eq:RC-definition} and after some manipulations, the renormalized current can be related to the bare one:
\begin{equation}
    J_{\mathrm{ren},\alpha} = J_{\mathrm{bare},\alpha} - \sum_{S} h^S_{\alpha \beta} V^S_\beta + J'_{\alpha}, \label{eq:fundamental}
\end{equation}
where the precise forms of the time-independent $h$-matrices and the residual current $\mathbf{J}'$ are provided in Appendix~\ref{app:bare-renorm-current}.
Adding a non-diffusive signal does not change the transport coefficient: the desired result will thus follow from the non-diffusivity of the signal $\Delta\mathbf{J} = \mathbf{J}_\mathrm{bare} - \mathbf{J}_\mathrm{ren} = \sum_{S} h^S \mathbf{V}^S - \mathbf{J}'$. In Appendix~\ref{app:bare-renorm-current} we show that the residual current $\mathbf{J}'$ can be neglected in the thermodynamic (TD) limit, leading to the relation:
\begin{equation}
    \frac{1}{V} \langle\Delta\mathbf{J}(t)\cdot\Delta\mathbf{J}(0)\rangle dt \sim 
        \sum_{S,S'} h^{S}_{\alpha \beta} h^{S'}_{\alpha \beta'} \frac{\langle V^S_\beta(t) V^{S'}_{\beta'}(0) \rangle}{V} . \label{eq:asimp}
\end{equation}
The expectation value $\frac{1}{V} \langle V^S_\beta(t) V^{S'}_{\beta'}(0) \rangle \sim \mathcal{O}(1)$, thus bringing a non-vanishing contribution in the TD limit. 
However, as already observed, in solids, amorphous materials and one-component molecular liquids the signals $V^S_\beta$ are non-diffusive and therefore every cross-correlation $\langle V^S_\beta(t) V^{S'}_{\beta'}(0) \rangle$ has a vanishing zero-frequency component, as a consequence of the lemma in Ref.~\citenum{Marcolongo2016}. Therefore, even if the signal $\Delta\mathbf{J}$ in general shows a non-zero autocorrelation function, its integral from zero to infinity has a vanishing value in the TD limit.

\section{Relation between decorrelation and renormalization}  \label{sec:comparison}
The two methodologies are related by the fact, proven in Appendix~\ref{app:equivalence}, that in the TD limit the coefficients $h$ of Eq.~\eqref{eq:fundamental} solve the same linear system that was derived by the decorrelation criterion, Eq.~\eqref{eq:decorr-LS}, with $\{\mathbf{Y}_n\}=\{\mathbf{V}^S\}$.
According to Eq.~\eqref{eq:fundamental}, since in the TD limit the residual current can be neglected, one has that  $\mathbf{J}_{\mathrm{ren}} \sim \mathbf{J}_{\mathrm{bare}} - \sum_{S} h^S \mathbf{V}^S$ and therefore the decorrelation becomes equivalent to the energy-flux renormalization. This result poses an application of both methods on solids grounds. The decorrelation technique, however, can also be performed with respect to other types of non-diffusive signals, such as the adiabatic electronic current $\mathbf{J}_\mathrm{el}$ \CORR{(defined in Appendix~\ref{sec:electronic-current}).}

We compare numerically the two techniques using the \emph{ab initio} energy currents of a-SiO$_2$ and reporting all the estimated thermal conductivities in Table~\ref{tab:kappa-results}. These values and their statistical errors have been obtained using the cepstral analysis technique, \CORR{which exploits the statistical properties of the power spectrum of the energy current, $S(\omega)$, to estimate the thermal conductivity in an efficient and straightforward way.\cite{Ercole2017,thermocepstrum} The parameters used in the analysis are reported in Appendix~\ref{app:details}.}
As we previously noted, the bare current $\mathbf{J}^\mathrm{DFT}_\mathrm{bare}$ leads to an overestimation of $\kappa$ (field ``DFT bare'' in the table) that can be corrected by applying the renormalization procedure (``DFT ren''). Equivalently, one could decorrelate $\mathbf{J}^\mathrm{DFT}_\mathrm{bare}$ from one particle current\footnote{Since $\mathbf{V}^\mathrm{Si}$ and $\mathbf{V}^\mathrm{O}$ are trivially proportional, due to the conservation of total momentum, the decorrelation of a current with respect to one or the other yields the same results.}, \emph{i.e.} $\mathbf{V}^\mathrm{Si}$ or $\mathbf{V}^\mathrm{O}$, thus obtaining a signal that gives a $\kappa$ compatible with the one obtained from $\mathbf{J}^\mathrm{DFT}_\mathrm{ren}$ (``DFT bare, VD''). 
We also notice that the power spectra of these two signals are almost identical (labeled ``DFT bare + VD'' and ``DFT ren'' in Fig.~\ref{fig:silica-spectra}), confirming the equivalence of the decorrelation and renormalization methods. As one may suspect, decorrelation with respect to the particle velocities has no effect when applied to $\mathbf{J}_\mathrm{ren}$, in which the sum of all the renormalized velocities is zero by construction (see Appendix~\ref{app:equivalence}, Eq.~\eqref{app:decorrelation-ren} for a formal justification). 

\CORR{Moreover, one can try to decorrelate the current from other non-diffusive signals, in order to further decrease the power of the noise. 
For example, if we decorrelate $\mathbf{J}^\mathrm{DFT}_\mathrm{bare, VD}$ or $\mathbf{J}^\mathrm{DFT}_\mathrm{ren}$ from the electronic current $\mathbf{J}_\mathrm{el}$ (see definition in Appendix~\ref{sec:electronic-current}),} we obtain a signal whose power is reduced by an additional factor of $\approx 3$, and that gives a compatible thermal conductivity (fields  ``DFT bare, VED'' and ``DFT ren, ED'', respectively).

\CORR{However, the particle current appears to be the largest source of inert signals in this system. In fact, if we were to decorrelate the bare current $\mathbf{J}^\mathrm{DFT}_\mathrm{bare}$ solely with respect to $\mathbf{J}_\mathrm{el}$, we would not obtain a correct value of $\kappa$ (``DFT bare, ED''), because the power of the spectrum would still be too large (\emph{i.e.} just a factor $\approx 3$ smaller than the power of $\mathbf{J}^\mathrm{DFT}_\mathrm{bare}$), thus compromising its analysis.}

\begin{figure}[!tb]
    \centering
    \subfigure[\label{fig:water-gk}]{\includegraphics[scale=1.0]{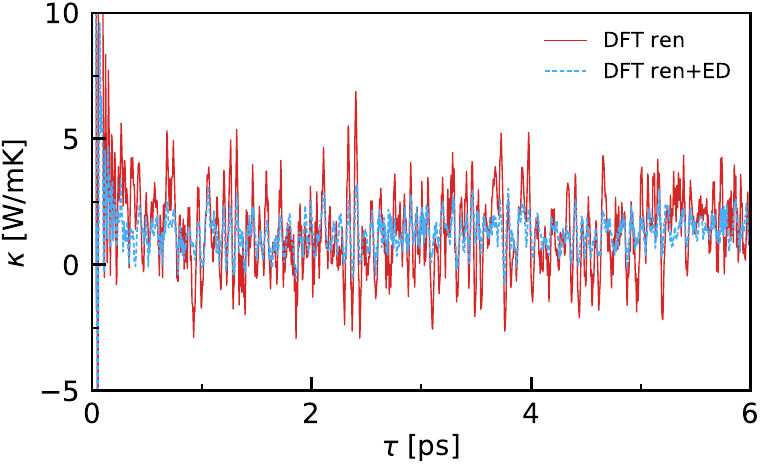}} \hfill
    \subfigure[\label{fig:water-spectra}]{\includegraphics[scale=1.0]{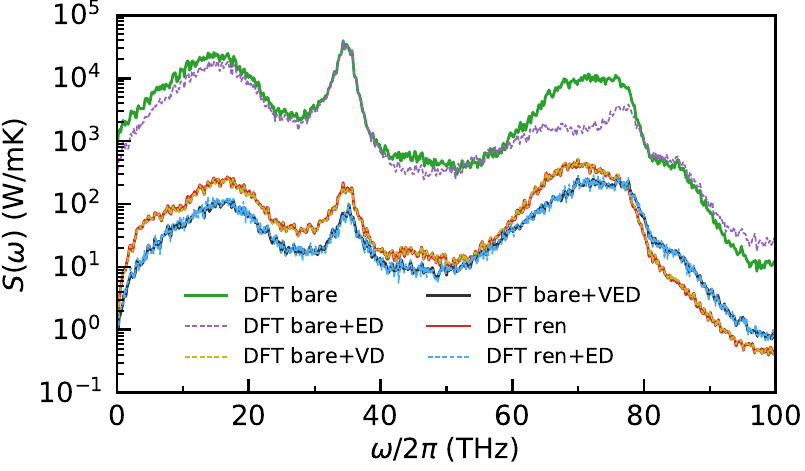}}
    \caption{
    Analysis of different definitions of the energy current of \emph{ab initio} water. The original $\mathbf{J}_\mathrm{bare}$ DFT definition has been renomalized, obtaining $\mathbf{J}_\mathrm{ren}$. Also reported are the same currents decorrelated with respect to the particle current of one species (VD), the electronic current (ED), or both (VED).
    (a) Thermal conductivity computed from the GK equation, Eq.~\eqref{eq:GK}, as a function of the upper integration limit. The result of the bare DFT current is out of scale and has been omitted. 
    (b) Power spectrum of the energy currents. The zero frequency corresponds to $\kappa$.}
    \label{fig:water}
\end{figure}

Finally, we performed the same analysis on \emph{ab initio} heavy water at ambient conditions. The simulation was performed with the same setting of Ref.~\citenum{Marcolongo2016}. 
Similarly to silica, an analysis of the bare current $\mathbf{J}^\mathrm{DFT}_\mathrm{bare}$ returns an unphysical value of $\kappa$, reported in Table~\ref{tab:kappa-results}. The renormalization procedure or the decorrelation with respect to a particle signal should be applied in order to obtain a physical value of $\kappa$, which is compatible with the one originally estimated by \citeauthor{Marcolongo2016}.\cite{Marcolongo2016}
Moreover, a decorrelation with respect to the electronic current $\mathbf{J}_\mathrm{el}$ can also be applied and gives compatible results. 
The GK integrals and the power spectra of the energy current for the different definitions are displayed in Fig.~\ref{fig:water}. 
\CORR{Ultimately, when dealing with DFT energy currents of solids or one-component liquids, the most effective and efficient strategy proved to be a velocity-renormalization of the energy-current time series followed by a decorrelation with respect to the electronic current. 
Alternatively, one may obtain similar results by considering the multi-component formalism.}

\section{Relation with the multi-component formalism}  \label{sec:multi-component}
\color{black} The variational framework presented in this work can be exploited to rederive, in an alternative way, the formulas used to evaluate the thermal conductivity of multi-component systems. Here we recall the basic result, following the presentation of Ref.~\citenum{Bertossa2019}. We consider a multi-component system characterized by $M$ linearly independent, conserved fluxes $\mathbf{J}_i$, $i=0,\dots M-1$, individually diffusive and where $\mathbf{J}_0$ is the energy flux.
We than define the matrix:
\begin{equation}
	\Lambda_{ij} \equiv \frac{1}{2} \int_{-\infty}^{\infty} \left\langle \mathbf{J}_i(t)\mathbf{J}_j(0) \right\rangle dt , \qquad  i,j=0,\dots M-1 \,.
\end{equation}
The thermal conductivity coefficient in the multi-component setting is then given by:
\begin{equation}
    \kappa = \frac{1}{3V k_B T^2} \frac{1}{\left[\Lambda^{-1}\right]_{00}} \,.
\label{eq:GK-multicomponent}
\end{equation}

We shall show that, by taking $\mathbf{Y} = \{\mathbf{J}_i\}_{i=1,\dots M-1}$ in Eq.~\eqref{eq:function} and applying our variational framework we can recover the formula of the multi-component theory. In particular, we first start with the bare energy flux $\mathbf{J}_0$ and decorrelate it with respect to the other conserved fluxes using an \emph{ad-hoc} engineered scalar product. 
We then evaluate the thermal conductivity by inserting the decorrelated $\mathbf{J}'_0$ into the single-component GK equation \eqref{eq:GK}. In this way we will recover Eq.~\eqref{eq:GK-multicomponent}, showing that the multi-component formula follows by dynamically decoupling the energy current from the other conserved fluxes.
\color{black}
Let us use in Eq.~\eqref{eq:function} the following \emph{Green-Kubo scalar product} between two generic fluxes,  $A$ and $B$:
\begin{equation}
	\left( A, B \right)_{\mathrm{GK},\omega} \equiv \frac{1}{2} \int_{-\infty}^{\infty} \left\langle A(t) B(0) \right\rangle e^{i\omega t} dt ,  \label{eq:gk-scalar-product}
\end{equation}
which is symmetric, bilinear and real. \color{black} As before, for vector observables like heat fluxes, the product $AB$ implies a scalar product as well \color{black} The symmetry property follows from the identity $\left\langle A(t) B(0) \right\rangle = \left\langle B(t) A(0) \right\rangle$, while the scalar product is real because $\left\langle A(-t) B(0) \right\rangle = \left\langle A(t) B(0) \right\rangle$. 
Both identities follow from Onsager's principle of microscopic reversibility \cite{Casimir1945} and reminding that fluxes are odd under time reversal. 
$(A,B)_\mathrm{GK,\omega}$ is also known as the cross-power spectrum, and $(A,A)_\mathrm{GK,\omega}$ is the power spectrum of $A$, which is always $\ge 0$ for stationary signals.
We note that the GK scalar product is well defined on the subspace generated by a set of signals that are odd under time reversal, and it depends parametrically on the chosen frequency $\omega$.
For small $\omega \ne 0$ we can assume the scalar product to be positive definite, whereas any non-diffusive signal has zero norm at $\omega=0$. 
Later we will discuss the limit $\omega \rightarrow 0$.

Let us define the (frequency-dependent) matrices: 
\begin{align}
	\Lambda_{ij}(\omega) &\equiv \left(\mathbf{J}_i, \mathbf{J}_j \right)_\mathrm{GK,\omega}, \quad & i,j=0,\dots M-1 ,\\
	\Sigma_{ij}(\omega) &\equiv \Lambda_{i,j}, \quad & i,j=1,\dots M-1 ,\\
	K_{i}(\omega) &\equiv \Lambda_{i,0} = \Lambda_{0,i}, \quad & i=1,\dots M-1 ,
\end{align}
where $\Lambda$ and $\Sigma$ are symmetric and $K$ is a column vector.
We now define:
$\mathbf{J}'_0 (\omega) = \mathbf{J}_0 - \sum_{i=1}^{M-1} \lambda_i(\omega) \mathbf{J}_i$.
For each $\omega$, the vector of coefficients $\{\lambda_i (\omega)\}_{i=1,\dots M-1}$ that minimizes Eq.~\eqref{eq:function} (\emph{i.e.} the power spectrum $\norm{ \mathbf{J}'_0(\omega)}_{\mathrm{GK},\omega}^2$) is then obtained by decorrelating the energy flux $\mathbf{J}_0$ with respect to the other particle fluxes and solving the linear system: $\lambda(\omega) = \Sigma(\omega)^{-1} K(\omega)$.
If we apply the (frequency-dependent) GK formula, Eq.~\eqref{eq:GK}, to the optimized energy flux $\mathbf{J}'_0 (\omega)$, we obtain:
\begin{align}
	\kappa(\omega) &= \frac{1}{3V k_B T^2} \left\lVert \mathbf{J}'_0(\omega) \right\rVert^2_\mathrm{GK,\omega} 
\end{align}

An explicit computation leads to:
\begin{align}
	\kappa(\omega) &= \frac{1}{3V k_B T^2} \left( \Lambda(\omega)_{00} - K(\omega)^\top \Sigma(\omega)^{-1} K(\omega) \right) \nonumber \\
	&= \frac{1}{3V k_B T^2} \frac{1}{\left[\Lambda(\omega)^{-1}\right]_{00}} ,  \label{eq:GK-omega-estimate}
\end{align}
where the last equality follows from writing the inverse of the $\Lambda$ matrix in a block form \cite{Bertossa2019}.
Interestingly, in the $\omega\rightarrow 0$ limit Eq.~\eqref{eq:GK-omega-estimate} becomes exactly the expression of the thermal conductivity for multi-component systems, which is usually derived from the Onsager relations by imposing the vanishing of all the mass fluxes \cite{Baroni2018,Bertossa2019}. 
The matrix $\Lambda_{ij}(\omega=0)= \left(\mathbf{J}_i, \mathbf{J}_j \right)_\mathrm{GK,\omega=0}$ can be assumed to be invertible in this multi-component setting.
Furthermore, $\kappa(\omega)$ is proportional to the so-called reduced spectrum, which was defined in Ref.~\citenum{Bertossa2019} and will always be lower than the power spectrum of both the original and the decorrelated signal, due to the present variational derivation.

In Ref.~\citenum{Bertossa2019} the formulas of the multi-component formalism were also applied as a tool to remove spurious signals in polyatomic liquids, by considering the energy flux $\mathbf{J}_0$ and a generic set of inert signals $\{\mathbf{J}_i\}_{i=1,..,M-1}$. 
\CORR{Inert signals are non-diffusive by definition, in contrast to the fluxes $\{\mathbf{J}_i\}_{i=1,..,M-1}$ considered in the multi-component case. As a consequence, in}
this case the only non-zero element of $\Lambda(\omega=0)$ is $\Lambda(\omega=0)_{00}=(\mathbf{J}_0,\mathbf{J}_0)_{\mathrm{GK},\omega=0}$ and the matrix is obviously non-invertible. Nevertheless, the limit $\omega \rightarrow 0$ can be taken after the matrix inversion, leading to a well defined scheme. The cepstral analysis technique described in Ref.~\citenum{Bertossa2019} provides a statistically correct way to obtain the $\omega \rightarrow 0$ limit after the matrix inversion.

\section{Conclusions}  \label{sec:conclusions}
In this work we presented a general framework, based on a variational principle, able to optimize a generic energy-flux time series by removing inert signals that do not contribute to the thermal conductivity. Our method is general and it can be generalized to the computation of other transport coefficients as well, whenever a non-diffusive signal can be identified. In the case of thermal transport, we highlighted why atomic binding contributions can pose serious convergence problems to Green-Kubo thermal conductivity simulations, especially in \emph{ab initio} frameworks and in polyatomic systems. We investigated numerically two solutions to this problem, that use the static cross-correlation as a scalar product between observables in our general framework. The first approach is based on the concept of decorrelation of the energy-current time series, and the second on renormalization of velocities. In the thermodynamic limit the two procedures are shown to be equivalent both from a theoretical and numerical point of view, when the decorrelation technique is used with the particle currents. The decorrelation technique can indeed be applied when a generic slowly-decaying signal makes a direct application of the GK formulas impossible or extremely expensive. The renormalization technique, instead, decorrelates the energy flux with respect to the particle current, but it is more straightforward to apply and can be used to detect whether binding energy contributions are correctly handled. As we demonstrated numerically, ignoring binding energy contributions can lead to wrong results, which cannot be detected by any standard statistical analysis of the heat currents. We therefore propose that both procedures should be performed to ensure that the simulation is not affected by convergence problems due to slowly-decaying signals. In our first-principles calculations we followed the heat-current definition of Ref.~\citenum{Marcolongo2016}, but other definitions can be affected by the same problems described in this work, when applied to polyatomic systems. 

Finally, we proved formally the equivalence between our general framework and the GK theory of heat transport in multi-component systems, by identifying a scalar product that formalizes the dynamical decoupling between the fluxes that are associated to the different transport mechanisms in such systems. We think that this work will help to interpret and analyze future applications of the GK theory to the computation of thermal conductivity, both using DFT or advanced force fields.

\medskip
\section{Acknowledgments}
LE and SB would like to thank Federico Grasselli for many valuable discussions. This work was partially funded by the EU through the \textsc{MaX} Centre of Excellence for supercomputing applications (Project No.~676598).

The authors declare no competing financial interests.

%%%%%%%%%%%%%%%%%%%%%%%%%%%%%%%%%%%%%%%%%%%%%%%%%%%%%%%%

\appendix

%ACTIVATE FOR ACHEMSO
%\appendixpage

\section{General relation between bare and renormalized currents}  \label{app:bare-renorm-current}
We prove here Eq.~\eqref{eq:fundamental} of the main text, showing the explicit form of the $h$-matrices and of $\mathbf{J}'$. 
We start from Eq.~\eqref{eq:generic} and denote with the an arrow $\overset{\mathrm{ren}}{\longrightarrow}$ the substitution of velocities with their renormalized values. We also introduce a convenient linear operator $\delta$, acting on a phase-space observable $X$ as $\delta X \equiv X - \langle  X \rangle$, which isolates the fluctuations.\\
Under renormalization, the term linear with the velocities transforms in the following way:
%ACTIVATE FOR ACHEMSO
%\begin{align}
%    \sum_i \epsilon^i_{\alpha \beta}(\{\mathbf{r}\}) v_{i,\beta} \;\overset{\mathrm{ren}}{\longrightarrow}\;
%        \sum_i \epsilon^i_{\alpha \beta}(\{\mathbf{r}\}) v_{i,\beta} 
%        - \sum_{S} \left[ \langle\varepsilon_{\alpha \beta}^S\rangle + \frac{1}{N_S} \sum_{i\in S} \delta \varepsilon^{i}_{\alpha \beta} \right] V^S_\beta, 
%    \label{eq:apx-one}
%\end{align}
%REMOVE FOR ACHEMSO
\begin{multline}
    \sum_i \epsilon^i_{\alpha \beta}(\{\mathbf{r}\}) v_{i,\beta} \;\overset{\mathrm{ren}}{\longrightarrow}\;
        \sum_i \epsilon^i_{\alpha \beta}(\{\mathbf{r}\}) v_{i,\beta} \\
        - \sum_{S} \left[ \langle\varepsilon_{\alpha \beta}^S\rangle + \frac{1}{N_S} \sum_{i\in S} \delta \varepsilon^{i}_{\alpha \beta} \right] V^S_\beta, %\left( \sum_{i'_A} v_{i'_A,\beta} \right).
    \label{eq:apx-one}
\end{multline}

where, with a slight abuse of notation, we called $\langle\varepsilon^{S}_{\alpha,\beta}\rangle$ the ensemble mean of $\langle  \varepsilon^{i}_{\alpha,\beta} \rangle$ over all atoms of species $S$, supposed to be equivalent. 
The kinetic term instead transforms like this:

%ACTIVATE FOR ACHEMSO
%\begin{align}
%    \sum_{i}\left(\frac{1}{2}m_{i}v_i^2\right)v_{i,\alpha} \;\overset{\mathrm{ren}}{\longrightarrow}\;
%        \sum_{i}\left(\frac{1}{2}m_{i}v_i^2\right)v_{i,\alpha} \\
%        - \sum_{S} \left[ \left(\sum_{i\in S} \frac{1}{2}m_{i}v_{i}^2\right) \frac{V^S_\alpha}{N_S} 
%        - m_S \left(\sum_{i\in S} v_{i,\alpha} v_{i,\beta}\right) \frac{V^S_\beta}{N_S} \right. 
%        \left.+ m_S V^S_\alpha \frac{(V^S)^2}{N^2_S} \right] % 
%    \label{eq:apx-two}
%\end{align}

%REMOVE FOR ACHEMSO
\begin{multline}
    \sum_{i}\left(\frac{1}{2}m_{i}v_i^2\right)v_{i,\alpha} \;\overset{\mathrm{ren}}{\longrightarrow}\;
        \sum_{i}\left(\frac{1}{2}m_{i}v_i^2\right)v_{i,\alpha} \\
        - \sum_{S} \left[ \left(\sum_{i\in S} \frac{1}{2}m_{i}v_{i}^2\right) \frac{V^S_\alpha}{N_S} %\left(\sum_{i'_A}  v_{i'_A,\alpha} \right)\nonumber \\
        - m_S \left(\sum_{i\in S} v_{i,\alpha} v_{i,\beta}\right) \frac{V^S_\beta}{N_S} \right.\\% \left( \sum_{i'_A} v_{i'_A,\beta} \right) \nonumber \\ 
        \left.+ m_S V^S_\alpha \frac{(V^S)^2}{N^2_S} \right] % \left(\sum_{i\in A} v_{i,\alpha}\right) \left( \sum_{i'_A} v_{i'_A,\beta} \right)  \left( \sum_{i''_A} v_{i''_A,\beta} \right).
    \label{eq:apx-two}
\end{multline}

The mean values are determined by the equipartition theorem, as:
\begin{eqnarray}
    \frac{1}{2}m_{i}v_i^2 &=& \frac{3}{2}k_B T + \frac{1}{2}m_{i} \delta(v_i^2) ,\\
    v_{i,\alpha} v_{i,\beta} &=& \frac{k_BT}{m_A} \delta_{\alpha \beta} +  \delta(v_{i,\alpha} v_{i,\beta}) ,
    \label{eq:apx-delta}
\end{eqnarray}

%ACTIVATE FOR ACHEMSO
%thus leading to the following substitutions in Eq.~\eqref{eq:apx-two}:
%\begin{align}
%    -\sum_{S} \left(\sum_{i\in S} \frac{1}{2}m_{i}v_{i}^2\right) \frac{V^S_\alpha}{N_S} = 
%    -\sum_{S} \left[ \frac{3}{2}k_B T + \frac{1}{2}m_S \frac{1}{N_S} \sum_{i\in S} \delta(v_{i}^2) \right] V^S_\alpha ,
%    \label{eq:apx-three}
%\end{align}
%and
%\begin{align}
%    -\sum_{S} m_S \left(\sum_{i\in S} v_{i,\alpha} v_{i,\beta}\right) \frac{V^S_\beta}{N_S} =
%    -\sum_{S} \left[ k_B T \delta_{\alpha,\beta} + m_S \frac{1}{N_S} \sum_{i\in S} \delta(v_{i,\alpha} v_{i,\beta}) \right] V^S_\beta .  
%    \label{eq:apx-four}
%\end{align}

%REMOVE FOR ACHEMSO
thus leading to the following substitutions in Eq.~\eqref{eq:apx-two}:
\begin{multline}
    -\sum_{S} \left(\sum_{i\in S} \frac{1}{2}m_{i}v_{i}^2\right) \frac{V^S_\alpha}{N_S} = \\%\left(\sum_{i'_A}  v_{i'_A,\alpha} \right)=\nonumber \\
    -\sum_{S} \left[ \frac{3}{2}k_B T + \frac{1}{2}m_S \frac{1}{N_S} \sum_{i\in S} \delta(v_{i}^2) \right] V^S_\alpha ,
    %-\sum_{A\in\mathbb{S}} \frac{1}{N_A} \left( \sum_{i_A} \delta\left(\frac{1}{2}m_{i}v_{i_A}^2\right) \right) \left(\sum_{i'_A}  v_{i'_A,\alpha} \right)\nonumber \\
    %-\sum_{A\in\mathbb{S}} \frac{3}{2}k_B  T \left(\sum_{i'_A}  v_{i'_A,\alpha} \right)
    \label{eq:apx-three}
\end{multline}
and
\begin{multline}
    -\sum_{S} m_S \left(\sum_{i\in S} v_{i,\alpha} v_{i,\beta}\right) \frac{V^S_\beta}{N_S} = \\% \left( \sum_{i'_A} v_{i'_A,\beta} \right) \nonumber \\
    %-\sum_{A\in\mathbb{S}} \frac{m_A}{N_A}\left(\sum_{i_A} v_{i_A,\alpha} v_{i_A,\beta} \right) \left( \sum_{i'_A} v_{i'_A,\beta} \right)= \nonumber \\
    -\sum_{S} \left[ k_B T \delta_{\alpha,\beta} + m_S \frac{1}{N_S} \sum_{i\in S} \delta(v_{i,\alpha} v_{i,\beta}) \right] V^S_\beta . %\left( \sum_{i'_A} v_{i'_A,\beta} \right) \nonumber \\
    %-\sum_{A\in\mathbb{S}} \left( \delta_{\alpha,\beta}k_B T \right) \left( \sum_{i'_A} v_{i'_A,\beta} \right). 
    \label{eq:apx-four}
\end{multline}
Combining Eqs.~\eqref{eq:apx-one}, \eqref{eq:apx-two}, \eqref{eq:apx-three}, and \eqref{eq:apx-four} we finally derive Eq.~\eqref{eq:fundamental} by defining the following $h$-matrices:
\begin{equation}
    h^{S}_{\alpha \beta} = \langle \varepsilon_{\alpha \beta}^S\rangle + \frac{5}{2}k_B T \delta_{\alpha \beta}. \label{eq:h-matrix}
\end{equation}
The residual current $\mathbf{J}'$ has the following form:
\begin{align}
    J'_\alpha = - \sum_{S} \left[
        \left(\frac{1}{N_S} \sum_{i\in S} \delta \epsilon^{i}_{\alpha \beta}\right)
        + m_S \left(\frac{1}{N_S} \sum_{i\in S} \delta(v_{i,\alpha}v_{i,\beta})\right) \right.\\
        \left. + \frac{1}{2} m_S \left(\frac{1}{N_S} \sum_{i\in S} \delta(v_{i}^2)\right) \delta_{\alpha \beta}
        - m_S \frac{(V^S)^2}{N^2_S} \delta_{\alpha \beta}
        \right] V^S_\beta .
        \label{eq:apx-five}
\end{align}
We now step back for a moment to define an $E$-signal as an observable which can be written as a sum of fluctuations $\delta f$, $E=\sum_{i} \delta f(\mathbf{r}_i, \mathbf{v}_i)$, \emph{i.e.} an extensive variable which can be written as a sum of local variables with zero mean. The precise form of the function $f$ depends on the particular $E$-signal and the sum over $i$ may be restricted to a certain atomic species. One can easily analyze the TD scaling of the autocorrelation functions formed by a finite product of $E$-signals. Expanding the summations and noticing that fluctuations among particles at large distances are uncorrelated the following result can be obtained:
\begin{equation}
    \langle E_1 \cdots E_n \rangle \sim \mathcal{O}(N^{\lfloor n/2 \rfloor}), \quad n \ge 2,
\label{eq:apx-six}
\end{equation}
where $\lfloor \cdot \rfloor$ is the floor function.\\
The residual current in Eq.~\eqref{eq:apx-five} has been  written explicitely as a finite sum of contributions, each one given
by a product of $E$-signals: $(E_1\times\cdots \times E_p)/N_S^q$, with $(p,q)=(2,1)$ or $(p,q)=(3,2)$, depending on the term considered.

 One can use this E-decomposition of $\mathbf{J}'$, Eq.~\eqref{eq:apx-five}, combined with the scaling of the individual $E$-signals, Eq.~\eqref{eq:apx-six}, to show by inspection that the terms neglected in Eq.~\eqref{eq:asimp}, i.e. terms of the form $\frac{1}{V} \langle V^S J' \rangle$ and $\frac{1}{V} \langle J' J' \rangle$, are indeed negligible in the TD limit. 

\section{Equivalence between renormalization and decorrelation}  \label{app:equivalence}
Let us consider only isotropic systems, such as $h^{S}_{\alpha \beta} \equiv h^{S} \delta_{\alpha \beta}$. 
In this section we use different brackets $\langle\cdot\rangle$ and $\langle\cdot\rangle_\mathrm{can}$ to indicate an equilibrium average in the microcanonical and the canonical ensemble, respectively. 
From the explicit form of the canonical Boltzmann distribution:
\begin{equation}
    P_\mathrm{can} (\{ \mathbf{r}, \mathbf{v}\}) \sim \exp \left[-\frac{U(\{ \mathbf{r}\})}{k_B T} \right] \prod_i  \exp \left[-\frac{m_i v_i^2}{2k_B T}\right],
\end{equation}
one can explicitly verify that each renormalized velocity $\mathbf{v}_i'= \mathbf{v}_i-\mathbf{W}^{S(i)}$ is independent of $\mathbf{V}^S$, for any species $S$. 
For example, given the linearity of the transformation and the Gaussian velocity distribution, it is sufficient to check that $\langle \mathbf{v}_i' \cdot \,\mathbf{V}^S \rangle_\mathrm{can} = 0$.
As a consequence $\langle \mathbf{J}_{\mathrm{ren}} \cdot \mathbf{V}^S\rangle_\mathrm{can} = 0$, the current $\mathbf{J}_{\mathrm{ren}}$ being sum of functions of $\mathbf{v}_i'$ and therefore uncorrelated with $\mathbf{V}^S$. 
This result holds exactly for every finite number of particles and not only in the TD limit. 

The standard formula relating the expectation values of fluctuations in different ensembles \cite{Lebowitz1967,Wallace2003,MarcolongoHaven} now reads:
\begin{align}
    \langle \mathbf{J}_{\mathrm{ren}} \cdot \mathbf{V}^S \rangle = \sum_{\alpha,\beta} \frac{\partial (\beta w_{\alpha})}{\partial M_{\beta}}
        \left( \frac{ \partial \langle \mathbf{J}_{\mathrm{ren}} \rangle_\mathrm{can}}{\partial (\beta w_{\alpha})}
        \frac{ \partial \langle \mathbf{V}^S \rangle_\mathrm{can}}{\partial (\beta w_{\beta})}   \right) + \\
    + \frac{\partial \beta }{\partial E} \left( \frac{ \partial \langle \mathbf{J}_{\mathrm{ren}} \rangle_\mathrm{can}}{\partial \beta} 
        \frac{ \partial \langle \mathbf{V}^S \rangle_\mathrm{can}}{\partial \beta } \right) + \smallO(N) \label{app:decorrelation-ren},
\end{align}
where $E$ and $\mathbf{M}$ are the total energy and momentum, whose conjugate quantities are $\beta$ and $\beta\mathbf{w}$, $\mathbf{w}$ being the mean centre
of mass velocity. 
The renormalized current has a zero expectation value for every temperature and is by construction independent of any global drift of the system.
Therefore all derivatives of $\mathbf{J}_{\mathrm{ren}}$ are zero and $\langle \mathbf{J}_{\mathrm{ren}} \cdot \mathbf{V}^S\rangle \sim \smallO(N)$. 
Let us exploit this result and calculate the scalar product of $\mathbf{J}_\mathrm{ren}$ and a generic $\mathbf{V}^{S'}$. 
Using Eq.~\eqref{eq:fundamental}, in the TD limit one finds the following formal set of relations satisfied by $h^S$:
\begin{equation}
    \left\langle \left( \mathbf{J}_\mathrm{bare} - \sum_{S} h^{S} \mathbf{V}^S \right) \cdot \mathbf{V}^{S'} \right\rangle = 0 , \quad \forall S' ,
\end{equation}
which is exactly the same linear system used for the decorrelation technique. As discussed in the text, this implies the equivalence, in the TD limit, between the decorrelation and renormalization techniques.

\section{Numerical check of theoretical thermodynamic scalings}
\label{app:dft-signals}
\begin{figure}[tb]
    \includegraphics[scale=1.0]{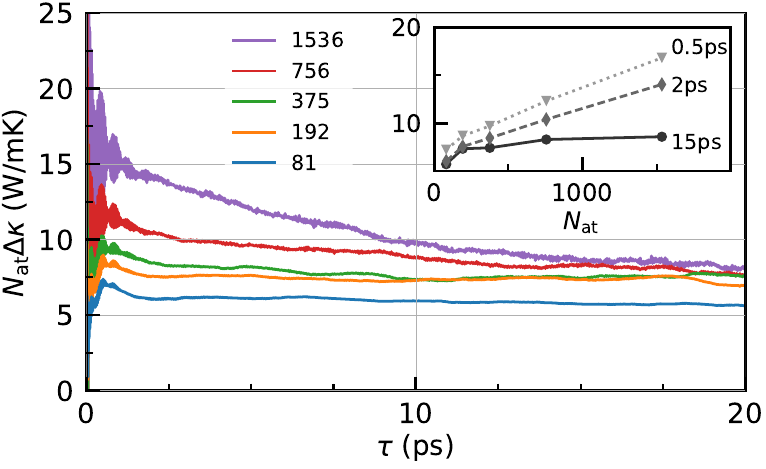}
    \caption{
    Behavior of $N_\mathrm{at} \Delta\kappa(\tau)$, defined in Eq.~\eqref{eq:delta-kappa}, for a flexible water model at $300\un{K}$. Each curve refers to a different number of particles $N_\mathrm{at}$, reported in the legend. Inset: $N_\mathrm{at} \Delta\kappa(\tau)$ at fixed $\tau$, as a function of $N_\mathrm{at}$. At short times the magnitude grows with $N_\mathrm{at}$, since $\Delta \kappa(\tau)$ tends to a finite limit; instead, the converged value for large $\tau$ remains stable and compatible with a $\mathcal{O}(1/N_\mathrm{at})$ decay of $\Delta \kappa(\tau)$.}
    \label{fig:NDelta}
\end{figure}

We can numerically check the TD scaling of $\Delta\mathbf{J}=\mathbf{J}_\mathrm{bare} - \mathbf{J}_\mathrm{ren}$ predicted by Eq.~\eqref{eq:asimp}, for the classical water model presented in the section \ref{sec:techniques}.
We focus our attention on the TD scaling for large number of atoms $N_\mathrm{at}$ of the function:
\begin{equation}
    N_\mathrm{at} \Delta \kappa(\tau) = \frac{N_\mathrm{at}}{3Vk_BT^2} \int_0^\tau \langle \Delta\mathbf{J}(t) \Delta\mathbf{J}(0)\rangle dt , \label{eq:delta-kappa}
\end{equation}
plotted in Fig.~\ref{fig:NDelta}.
From the arguments presented in section  \ref{sec:techniques}, two different TD scaling regimes are expected and observed as a function of $\tau$:
at short times $\Delta \kappa(\tau)$ tends to a finite value (in the TD limit) given by Eq.~\eqref{eq:asimp}, thus $N_\mathrm{at} \Delta \kappa(\tau)$ diverges linearly with $N_\mathrm{at}$; 
at long time lags the contribution given by Eq.~\eqref{eq:asimp} integrates to zero and we observe that $N_\mathrm{at} \Delta \kappa(\tau)$ tends to a finite limit, hence $\Delta \kappa(\tau) \sim \mathcal{O}(1/N_\mathrm{at})$.

\color{black}
\section{Computational details}  \label{app:details}

\paragraph{Classical Water - Simulation details}
Classical simulations of water have been performed using \textsc{LAMMPS}\cite{LAMMPS}, considering a flexible model of water\cite{water-FF}, in a cubic box with an edge of $18.6\un{\angstrom}$ containing $216$ molecules, obtaining the experimental density. The integration time step was $0.5\un{fs}$. The system was thermalized in the NVT ensemble at $300\un{K}$ for $200\un{ps}$ using the stochastic velocity rescaling thermostat\cite{Bussi2007} with a coupling time of $200\un{fs}$, followed by $50\un{ps}$ of equilibration in the NVE ensemble.
Subsequently, the system has been evolved in the NVE ensemble for $10\un{ns}$, and the heat flux has been computed every $1\un{fs}$ using the classical definition, Eq.~\eqref{eq:classical-heat-flux}.

\paragraph{Silica - Simulation details}
The classical trajectory of a-SiO$_2$ has been generated using \textsc{LAMMPS}\cite{LAMMPS} and the BKS force field\cite{Silica-BKS-1990}. A previously generated 72-atom sample of a-SiO$_2$ was considered, in a cubic box with an edge of $20.3\un{\angstrom}$. The integration time step was $1.0\un{fs}$. The system was thermalized in the NVT ensemble at $300\un{K}$ for $500\un{ps}$, followed by $100\un{ps}$ of equilibration in the NVE ensemble.
Subsequently, the system has been evolved in the NVE ensemble for $1\un{ns}$, while the heat flux was computed every $1\un{fs}$ using the classical definition, Eq.~\eqref{eq:classical-heat-flux}. $100\un{ps}$ of this trajectory were used for the \emph{ab initio} calculation of the heat flux. 
To estimate the thermal conductivity, we used cepstral analysis\cite{Ercole2017} with $f^*\approx 17\un{THz}$. We verified that the value of $\kappa$ does not vary with respect to this choice.

\paragraph{\emph{Ab initio} Water - Simulation details}
The simulation setting is the same as Ref.~\citenum{Marcolongo2016}. The trajectory length analysed was $90\un{ps}$. The parameters used in cepstral analysis are $f^*\approx 9\un{THz}$, and in this case the $P^*$ value was multiplied by a factor $1.5$ to reduce any bias possibly due to the fast variation of the spectrum at frequency close to zero.

\paragraph{Green-Kubo integral}
The simplest way to estimate the thermal conductivity $\kappa$ consists in a direct calculation of the GK equation \eqref{eq:GK} as a function of the upper limit of integration. We have that $\kappa = \lim_{\tau\rightarrow\infty} \kappa(\tau)$, where $\kappa(\tau) = \frac{1}{3Vk_B T^2} \int_0^\tau \! \left\langle \mathbf{J}(t) \cdot \mathbf{J}(0) \right\rangle dt$. The autocorrelation function $\left\langle \mathbf{J}(t) \cdot \mathbf{J}(0) \right\rangle$ can be evaluated numerically from a discretized-time series as an average over many time origins:\cite{Allen1989}
\begin{equation}
    \left\langle \mathbf{J}(t) \cdot \mathbf{J}(0) \right\rangle = \frac{1}{t_\mathrm{max}} \sum_{\alpha=1}^3 \sum_{t_0=0}^{t_\mathrm{max}} J^\alpha(t_0 + t) J^\alpha(t_0) , \label{eq:J-ACF}
\end{equation}
where $\alpha=1,2,3$ indicate the carthesian components, and we assume $t_\mathrm{max}+t \leq t_\mathrm{run}$. An estimate of the statistical error of Eq.~\eqref{eq:J-ACF} and of the estimator $\kappa(\tau)$ can be obtained by block analysis,\cite{Allen1989,Frenkel2002} however the convergence value of $\kappa(\tau)$ for $\tau\rightarrow\infty$ is generally difficult to identify and more advanced techniques such as cepstral analysis are needed.

\paragraph{Classical heat flux}
In a system of $N$ atoms interacting through a classical force field $U(\mathbf{r}_1, \mathbf{r}_2, \cdots \mathbf{r}_N)$ the ``standard'' definition of the energy density is written in terms of local atomic energies as:
\begin{align}
    & \epsilon(\mathbf{r}, \Gamma_t) = \sum_i \delta(\mathbf{r} - \mathbf{r}_i) \epsilon_i(\Gamma_t) \,, \\
    & \epsilon_i(\Gamma_t) = \frac{1}{2} m_i v_i^2 + U_i(\{\mathbf{r}\}) + \bar\epsilon^{S(i)} \,,
\end{align}
where $U_i$ are atomic potential energies whose sum is the total potential energy of the system, and $\bar\epsilon^{S(i)}$ is a species-dependent formation energy that is usually set to zero.
The energy flux can be written as:
\begin{equation}
    \mathbf{J}(t) = \sum_i \mathbf{v}_i \epsilon_i + \sum_{i,j} (\mathbf{r}_i - \mathbf{r}_j) \mathbf{f}_{ij} \cdot \mathbf{v}_i \,, \label{eq:classical-heat-flux}
\end{equation}
where $\mathbf{f}_{ij} = -\frac{\partial U_j}{\partial \mathbf{r}_i}$ is the contribution of the $j$-th atom to the force acting on the $i$-th atom, $\sum_j \mathbf{f}_{ij} = \mathbf{f}_i$, and $\mathbf{f}_{ij} = - \mathbf{f}_{ji}$. 
In the case of two-body potentials, such as the ones considered in this paper, the common choice consists in splitting the potential energy evenly between the two interacting atoms, that is: $U_j = \frac{1}{2}\sum_i U(\mathbf{r}_i-\mathbf{r}_j)$ and $\mathbf{f}_{ij} = -\frac{1}{2}\nabla_{\mathbf{r}_i} U(\mathbf{r}_i - \mathbf{r}_j)$. This is of course an arbitrary choice: the gauge invariance principle ensures that any other choice such that $\sum_i U_i = U$ leads to the same thermal conductivity.\cite{Ercole2016}

For two-body force fields, Eq.~\eqref{eq:classical-heat-flux} is implemented in \textsc{LAMMPS} and can be computed using the \texttt{compute heat/flux} command.

\paragraph{First-principles heat flux}
We report here for completeness the formulas used for the evaluation of the first-principles (DFT) heat flux. The bare current is given by the sum of the following components:
\begin{align}
  \mathbf{J}_{KS} 
  & =\sum_{v}
    \left (
    \langle\varphi_{v}| \mathbf{r}\hat{H}_{KS}|
    \dot{\varphi}_{v}\rangle + 
    \varepsilon_v \langle\dot{\varphi}_{v}|
                    \mathbf{r} |
    \varphi_{v}\rangle \right), \label{eq:J_KS}  \\
  \mathbf{J}_{H} & =\frac{1}{4\pi} \int\dot{u}_{H}(\mathbf{r})
                   \nabla u_{H}(\mathbf{r}) d\mathbf{r},\label{eq:J_H}\\
  \mathbf{J}'_{0} & =\sum_{i} \sum_v \left\langle \varphi_{v}
                   \left|(\mathbf{r}-\mathbf{r}_{i}) \left(\mathbf{v}_{i}
                   \cdot
                   \nabla_{\mathbf{r}_i} \hat{v}_0
                   \right)\right|\varphi_{v} \right\rangle ,\label{eq:J_0}\\
  \mathbf{J}_{0} & =\sum_{i} \Bigl [ \mathbf{v}_{i} e^0_i
                   + \sum_{j\ne i}
                   (\mathbf{r}_{i}- \mathbf{r}_{j})
                   \left(\mathbf{v}_{j} \cdot
                   \nabla_{\mathbf{r}_j} w_i
                   \right)\Bigr ], \label{eq:J_n} \\
  \mathbf{J}_{XC} &=\begin{cases} 0 & \text{(LDA)} \\
      -\int\rho(\mathbf{r})\dot{\rho}(\mathbf{r}) \partial\epsilon_{GGA}
      (\mathbf{r})d\mathbf{r} & \text{(GGA)}, \label{eq:J_XC} \end{cases}
\end{align}
where $\mathbf{r}_{i}$ and $\mathbf{v}_{i}$ are the positions and velocities of the $i$-atom. The electronic degrees of freedom are described by the wavefunctions $\varphi_{v}$. The electron charge is assumed to be one and the remaining symbols are instead defined as:
\begin{itemize}
   \setlength\itemsep{0pt}
   \item  $e^0_i$ : ionic energy, equal to $\frac{1}{2}m_{i}v_{i}^{2} + w_{i}$ ;
   \item $\epsilon_{XC}$ : local XC energy per particle, defined by the relation: $E_{XC}= \int \epsilon_{XC} [\rho](\mathbf{r})\rho(\mathbf{r})d\mathbf{r}$.  ``LDA'' and
  ``GGA'' in Eq.~\eqref{eq:J_XC} indicate the local-density and generalized-gradient approximations to the XC energy functional ;
   \item  $\epsilon_{v}$ : eletronic eigenvalues ;
   \item $\hat{H}_{KS}$ : instantaneous Kohn-Sham (KS) Hamiltonian ;
   \item $m_i$ : atomic mass ;
   \item $\mathbf{r}$ : multiplicative position operator ;
   \item $\rho(\mathbf{r})$ : ground-state electron-density distribution ;
   \item $u_{H}$ and $u_{XC}$ : Hartree and exchange-correlation (XC) potentials ;
   \item $\hat u_0$ : ionic (pseudo-) potential acting on the electrons ;
   \item $w_{i}$ : electrostatic energy, equal  to $\frac{1}{2}\sum_{j\ne i}\frac{Z_{i}Z_{j}}{|\mathbf{r}_{i}-\mathbf{r}_{j}|}$ ;    
   \item $Z_i$ : atomic charge ;
   \item $\partial\epsilon_{GGA}$ : derivative of the GGA XC local energy per particle with respect to density gradients ,
   \item $\nabla$ : gradient with respect to the spatial coordinate $\mathbf{r}$ ;
   \item $\nabla_{\mathbf{r}_i}$ : gradient with respect to the atomic position $\mathbf{r}_i$ ;
   \item $\langle \: \rangle$ : standard scalar product between wavefunctions ;
   \item $\dot{[\:]}$ :  derivative with respect to time ;
\end{itemize}
More technical details on a possible implementation can be found in the references \cite{Marcolongo2014,Marcolongo2016}.

Finally, even if we reported the precise expressions used in this work, we note that the decorrelation and renormalization techniques are general and do not depend on the specific definition of the bare heat flux. As discussed in the main text, the decorrelated current is evaluated by solving the linear system of Eq.~\eqref{eq:decorr-LS}, and plugging in the resulting $\{\lambda_n\}$ coefficients into Eq.~\eqref{eq:new-current}. The renormalized currents are instead evaluated using the same expression of the bare ones, but replacing velocities with their renormalized values.
\color{black}

\color{black}
\section{Electronic current}  \label{sec:electronic-current}
The adiabatic electronic flux $\mathbf{J}_\mathrm{el}$ can be evaluated as:\cite{Thouless1983}
\begin{equation}
    \mathbf J_\mathrm{el} = 2 \mathfrak{Re} \sum_{v} \left\langle \varphi_{v} \middle| \mathbf{r} |\dot \varphi_{v} \right\rangle ,
    \label{eq:electronic-flux}
\end{equation}
following the same notation of Appendix~\ref{app:details}. This expression can be derived from the conservation equation for the electronic density  $\rho$:
\begin{equation}
    \nabla\cdot \mathbf{j}_\mathrm{el}(\mathbf{r},t) =- \dot{\rho}(\mathbf{r},t) ,
\end{equation}
where  $\mathbf{j}_\mathrm{el}(\mathbf{r})$ is the electronic current density. The flux is then defined as $\mathbf J_\mathrm{el} = \int \mathbf j_\mathrm{el}(\mathbf{r}) d\mathbf{r} \sim \int  \dot{\rho}(\mathbf{r},t) \,\mathbf{r}\, d\mathbf{r}$, where boundary terms can be neglected \cite{TSun,Baroni2018}. Inserting in the last expression the definition of $\rho$ in terms of the electronic wavefunctions, one obtains Eq.~\eqref{eq:electronic-flux}, which is an expression well defined under periodic boundary conditions.\cite{Marcolongo2014} We note that the same current can be alternatively evaluated in terms of the Born-effective charges \cite{TSun}. 

The electronic current is the difference between the total charge current (defined as the atom's Born charge times its velocity and summed over the atoms) and its ionic component (here defined considering the nucleus plus the valence electrons). 
Analogously to the particle current, in the electrically insulating systems considered in this work the electronic current $\mathbf{J}_\mathrm{el}$ is a non-diffusive signal (because the difference between the electronic and the particle current is itself a non-diffusive signal).\cite{Marcolongo2016,Baroni2018,Grasselli2019}

\color{black}

\bibliography{renorm_biblio.bib}

\end{document}